\documentstyle[revtex,aps,prl]{revtex}

\tighten
\begin{document}

\draft

\title{Dynamics of Apparent and Event Horizons}

\author{Peter Anninos${}^{(1)}$, David Bernstein${}^{(1,*)}$,
Steven Brandt${}^{(1,3)}$, Joseph Libson${}^{(1,3)}$, Joan
Mass\'o${}^{(1,2)}$, Edward Seidel${}^{(1,3)}$, Larry
Smarr${}^{(1,3)}$, Wai-Mo Suen${}^{(4)}$, Paul Walker${}^{(1,3)}$}

\address{
$^{(1)}$ National Center for Supercomputing Applications,
Beckman Institute, 405 N. Mathews Ave., Urbana, IL, 61801 \\
$^{(2)}$ Departament de F\'{\i}sica, Universitat de les Illes Balears,
   E-07071 Palma de Mallorca, Spain. \\
${}^{(3)}$ Department of Physics,
University of Illinois, Urbana, IL 61801 \\
$^{(4)}$ McDonnell Center for the Space Sciences, Department of Physics,
Washington University, St. Louis, Missouri, 63130 \\
${}^{(*)}$ Present address: Department of Mathematics, Statistics
and Computing Science, \\
University of New England, Armidale, NSW 2351, Australia
}
\date{\today}

\maketitle

\begin{abstract}
The dynamics of apparent and event horizons of various black hole
spacetimes, including those containing distorted, rotating and
colliding black holes, are studied.  We have developed a powerful and
efficient new method for locating the event horizon, making possible
the study of both types of horizons in numerical relativity. We show
that both the event and apparent horizons, in all dynamical black hole
spacetimes studied, oscillate with the quasinormal frequency.
\end{abstract}


\narrowtext

\paragraph*{Introduction.}
Black holes are among the most fascinating predictions in the theory
of General Relativity.  During the past 20 years there have been
intense research efforts on black holes and their effect on the
astrophysical environment.  With the exciting possibility of detecting
gravitational wave signals from black holes by the gravitational wave
observatories under construction(LIGO and VIRGO~\cite{LIGO3}), we have
seen an ever increasing surge of interest.  The black hole events
likely to be observed by the gravitational wave observatories involve
highly dynamical black holes, e.g., two black holes in collision.  The
most powerful tool in studying such highly dynamical and intrinsically
non-linear events is probably numerical treatment.  In recent years,
there has been significant progress in numerical relativity in this
direction (See, e.g.,~\cite{Bona93}). In particular, long evolutions
of highly dynamical black hole spacetimes are now
possible~\cite{Abrahams92a,Anninos93b}, opening up the opportunity of
many interesting studies.

The essential characteristics of a black hole in relativity are its
horizons, in particular, the apparent horizon (AH) and the event
horizon (EH).  Singularity theorems~\cite{Hawking73a} link the
formation of black hole singularities to the formation of the AH.  The
membrane paradigm~\cite{Thorne86} characterizes black holes by the
properties of its EH, which is regarded as a 2D membrane
living in a 3D space, evolving in time and endowed with
many everyday physical properties like viscosity, conductivity,
entropy, etc.  We believe such a point of view is a powerful tool in
providing new insight into the numerical studies of black holes.

The apparent horizon (AH) at each instant of time is generically a
spacelike surface defined as the outermost trapped surface of a region
in space, whereas the event horizon (EH) is a null surface defined as
the boundary of the causal past of the future null infinity (For a
rigorous description, see, e.g.~\cite{Hawking73a}).  The AH is a local
object, readily obtained in a numerical evolution by searching a given
time slice for closed surfaces whose outgoing null rays have zero
expansion.  In contrast, the EH is a globally defined object which
makes it harder to find in numerical studies.  Its location, and even
its existence cannot be determined without knowledge of the complete
four geometry of the spacetime.

Despite the formal distinctions between the two kinds of horizons,
there are intriguing relations between them.  For example, the cosmic
censorship conjecture~\cite{Penrose69} (together with the singularity
theorems) suggests that the formation of an AH is usually accompanied
by the formation of an EH, and for all stationary
spacetimes~\cite{Carter79} the AH and EH coincide with each other.
How far can these similarities be extended?  Can the shape and surface
area of the two horizon surfaces be very different?  Can some
membrane-like~\cite{Thorne86} properties be associated with the AH?
These are some of the many interesting questions concerning the
properties of apparent and event horizons that we touch upon in this
paper and shall investigate further in future publications.

The central development that makes possible our study of horizon
dynamics is that we can now find both the AH and the EH in numerical
relativity.  In particular, we have developed a powerful and efficient
method to locate the EH in dynamic black hole spacetimes.  In this
paper, we outline this method and apply it to study and compare the
dynamical evolutions of the AH and EH for various spacetimes,
including single black holes interacting with external gravitational
fields and black holes in collision.  In particular we show that the
EH and the AH oscillate with the same frequency (the quasinormal
frequency as determined using black hole perturbation
theory~\cite{Chandrasekhar83}), even though the early stages of the
spacetime dynamics are highly nonlinear in the cases studied.

\paragraph*{Finding the Horizons.}
It is by now a routine exercise to find apparent horizons in numerical
relativity (see, e.g., Refs.~\cite{Cook90a}) and we will
not discuss AH methods further in this paper.  The actual EH of a
black hole can in principle be found by tracing the path of null rays
through time.  Outward going light rays emitted just outside the EH
will diverge away from it, escaping to infinity, and those emitted
just inside the EH will fall away from it, towards the singularity.
In a numerical integration it is difficult to follow accurately the
evolution of an actual null horizon generator forward in time, as
small numerical errors cause the ray to drift and diverge rapidly from
the true EH.  It is a physically unstable process.  But we can
actually use this property to our advantage by considering the
time-reversed problem.  Any outward going photon that begins near the
EH will rapidly be {\em attracted} to the horizon if integrated {\em
backward} in time. In integrating backwards in time, the initial guess
of the photon does not need to be very precise as it will converge to
the correct trajectory after only a short time. Therefore, we can
integrate backward starting from a position we expect to be near the
EH, e.g., the location of the AH at a time when the black hole has
settled down to an approximate stationary state after some dynamical
evolution.

In principle one could consider finding the EH by tracing a collection
of photons backwards in time through the numerically generated
spacetime. This necessitates integrating the geodesic equation for
each photon.  However, because the geodesic equation requires taking
derivatives of the metric functions, this procedure is extremely
sensitive to inaccuracies in the numerically generated metric. Also
(and more importantly) although photons will be attracted in the
normal direction to the EH, there is no such attractive property in
the tangential direction.  As the geodesic equation is 2nd order, the
initial directions of the photon must be specified.  The trajectory is
sensitive to the initial choice in the tangential direction, and it
may further drift tangentially due to inaccuracy in integration.  An
initial surface of photons is not guaranteed to be surface forming
after integration due predominantly to such tangential drifts.

Rather than independently tracking all individual photons starting on
a surface, we follow the entire null surface itself.  A null generator
of the null surface is guaranteed to satisfy the geodesic
equation~\cite{Carter79}.  A null surface defined by $f(t,x^i)=0$
satisfies the condition
\begin{equation}
g^{\mu\nu} \partial_{\mu} f \partial_{\nu} f = 0.
\label{nullsurface}
\end{equation}
Hence the evolution of the surface can be obtained by a simple
integration,
\begin{equation}
\partial_t f = \frac{ - g^{ti} \partial_i f -
\sqrt{(g^{ti}\partial_i f)^2 - g^{tt} g^{ij} \partial_i f \partial_j f}
}{g^{tt}}
\label{evolve}
\end{equation}
Notice that this equation contains only derivatives of the surface and
{\em not} of the metric components themselves and is therefore less
susceptible to the numerical inaccuracies present in the metric data.

The advantages to integrating an entire surface include: ({\it i})
Tangential drifting is not a source of error, because the only
direction that a surface can move is normal to itself; Once the
surface becomes the EH, it cannot drift away from it.  ({\it ii})
Unlike integrating null geodesics, this method is guaranteed to be
surface forming. ({\it iii}) It is simpler and more accurate than
evolving individual photons.

Using this method we are able to trace accurately the entire history
of the EH in a short period of time.  It takes just a few minutes to
trace the EH on a computer workstation for an axisymmetric spacetime
representing a black hole interacting with a gravitational wave (the
first case detailed below) resolved on a
grid of 200 radial by 53 angular zones and evolved to $t = 75M$ (where
$M$ is the mass of the black hole). We contrast our backward surface
method with another method~\cite{Hughes94a} that uses forward
integration of individual photons to find the EH.

\paragraph*{Horizons of Black Hole Spacetimes.}
The first case we discuss consists of a non-rotating black hole
surrounded by an axisymmetric gravitational wave
initially at a finite distance away from the hole.  The system was
evolved with a code described in Refs.~\cite{Bernstein93b,Anninos93c}.
The black hole becomes distorted as the incoming wave hits. In time,
it settles down and returns to a Schwarzschild hole with a larger
mass.  Fig.~(\ref{fig1}) shows the areas of the horizons vs. time.
Six different integrations of the EH starting at different places are
shown.  In one case the AH was used as an initial surface for the
integration.  Because the AH is {\em inside} the correct location of
the EH, the surface expands outward as it is attracted to the correct
location.  In other cases, surfaces larger or smaller than the AH are
chosen as initial guesses.  Note that in all cases the surfaces are
attracted to the {\em same} surface in precise detail, as they should
be. The insert shows an expanded view of the early time.  All surfaces
computed are shown, but they are completely indistinguishable in spite
of their extremely different starting positions, clearly showing the
power and stability of this method.  At $t=0$ the AH and EH
practically coincide with each other.  Then the EH foresees the coming
of the wave and expands.  As the wave is falling in, after about
$t=15M$, the AH starts to expand and catch up. The behavior of the AH
and EH are exactly as expected. (The area curves exaggerate the effect
of very small differences in coordinate location of the horizons, as
they are located in a region of the spacetime where the metric
functions have very steep gradients.  The numerically generated data
of the spacetimes do not generally resolve these gradients accurately,
leading to the spurious growing of the horizon area at late
time~\cite{seidel92a}.)  We have also studied the EH and AH of
distorted, rotating black holes.  We find the dynamical evolution of
those horizons to have similar properties.

In Fig.~(\ref{fig2}) we show a geometric embedding of the coalescing
horizons for the head-on collision of two black holes, as discussed in
Ref.~\cite{Anninos93b,Anninos94a}.  The embedding, which preserves the
proper surface areas of the horizons, shows not just the topology but
also the geometric properties of the horizon.  Although such a picture
of the embedding is familiar, this is the first time it has actually
been computed.  There have been a number of attempts to estimate the
critical separation parameter $\mu$ beyond which these initial data
sets contain two separate black holes~\cite{Smarr79}, and with our
method we can now say that for $\mu$ greater than about 1.8 there are
two holes.  The horizon shown in Fig.~(\ref{fig2}) corresponds to $\mu
= 2.2$, i.e., an initial proper separation of $8.92M$, where $M$ is
the mass of each hole.  We see that initially the two black hole
horizons are separated.  They coalesce at about $t=6M$, forming a
single large black hole.  Various properties of the horizons in the
process of coalescence will be analyzed elsewhere.

The ability to determine the AH and EH for dynamical holes opens up
the possibility for the first time of using the horizons as a tool to
study black hole physics in numerical relativity.  As a first example
of this, we compute the ratio of the polar to equatorial
circumference, $C_p/C_e$, of both the AH and EH for the first case
discussed above.  In Fig.~(\ref{fig3}) we show this ratio as a
function of time for both the AH and the EH.  [Note that the horizon
begins and ends as a sphere, as its gaussian curvature is constant to
within 1 part in $10^6$, which shows clearly that the hole was
``Schwarzschild'' at both times.  This also provides a stringent test
for our event horizon finder, since it must trace a surface backward
in time as it undergoes a period of distortion, and then return to a
sphere.]  We see that the AH and the EH oscillate in precisely the
same manner, despite the fact that one is a null surface while the
other is spacelike.  These oscillations of the horizons are caused by
the ``quasinormal mode ringing'' of the gravitational waves generated
by the potential barrier in the black hole spacetime.  As the waves
leak out to infinity and down the hole, those going down cause the
horizons to oscillate.  In the membrane point of view, the
oscillations are dissipated into viscous heating of the horizon
membrane, causing the horizon surface area to increase during the
oscillations, as we have seen in our calculations.  We show a fit of
the EH oscillation to the two lowest $\ell = 2$ quasinormal modes of
the black hole as determined in linear perturbation theory (the
initial incoming Brill wave is predominantly in the $\ell = 2$ mode).
The fit is remarkable, showing conclusively that both the AH and EH
oscillate at the natural frequency of the black hole.  (The horizons
oscillate with the quasinormal frequency in coordinate time, as they
should.)  We observe similar oscillations in both the distorted
rotating black holes and the black hole collisions, establishing the
generic nature of these results. This oscillation is also just visible
in the horizon embedding diagram (Fig.~(\ref{fig2})) for the two black
hole collision after the holes have coalesced, where about one
wavelength of the quasinormal mode is shown.

We have also found that for highly distorted holes, both AH and EH can
have $C_p/C_e$ very different from 1, making some circumference of the
horizon rather bigger than $4 \pi M$ at times.  The details of this and
its possible implication on the hoop conjecture (vacuum
version~\cite{Flanagan91a}) will be discussed elsewhere.

\paragraph*{Conclusions.}
We have developed a powerful method for finding black hole event
horizons in dynamic spacetimes based on the ideas of ({\it i})
backward integration and ({\it ii}) integrating the entire null
surface.  This opens up the possibility of studying the dynamics of
event horizons in numerical relativity.  We studied and compared the
behavior of both the apparent and event horizons for various dynamical
spacetimes.  In all cases studied, we show that the event and apparent
horizons oscillate with the quasinormal frequency of the black hole.

Our method can find event horizons without knowledge of the apparent
horizon, so it should be a useful tool for analyzing spacetimes even
in cases where the apparent horizon cannot be found (e.g., if the time
slicing does not intersect the apparent horizon).  Its impact on the
numerical investigations of the cosmic censorship conjecture and the
hoop conjecture could prove interesting.

We are very grateful to Scott Hughes, Charles R. Keeton II, Stuart
Shapiro, Saul Teukolsky, and Kevin Walsh for allowing us to use the
Cornell code for making comparisons with our code.  We also thank
S.H., C.R.K., and K.W., who, together with one of us (P.W.), developed
some of the I/O routines used in the present version of our
code. J.M. acknowledges a Fellowship (P.F.P.I.) from Ministerio de
Educaci\'on y Ciencia of Spain. This research is supported by the
NCSA, the Pittsburgh Supercomputing Center, and NSF grants Nos.
PHY91-16682 and ASC93-18152.



\begin{figure}
\caption{The area of the event horizon is traced through time for different
initial surfaces (solid lines), and compared to the area of the
apparent horizon (dashed line).  The attracting nature of the event
horizon is dramatic, as all of our backward surface integrations trace
the same path, although they start from very different initial
locations.  The insert shows an expanded view of the early time
results.  {\em All} surface integrations are shown, and are completely
indistinguishable.}
\label{fig1}
\end{figure}

\begin{figure}
\caption{The geometric embedding of the event horizon for two black holes
colliding head on is shown.  The $z$ coordinate marks the symmetry
axis, and $t$ is the coordinate time.  Initially the two holes are
separate, and by about $t = 6M$ they coalesce into a single black
hole.  The distance between the initial horizons is arbitrary in this
embedding space, but is based on the proper distance between the
holes.}
\label{fig2}
\end{figure}

\begin{figure}
\caption{The ratio $C_p/C_e$ of the polar circumference to the equatorial
circumference of both the event horizon (solid line) and apparent
horizon (dashed line) is shown versus coordinate time $t$ for a black
hole which is hit by a gravitational wave.  The dot-dashed line shows
the fit of the two lowest $\ell=2$ quasinormal modes to the event
horizon.  The hole oscillates at its quasinormal frequency when hit by
the wave.}
\label{fig3}
\end{figure}


\begin{thebibliography}{10}

\bibitem{LIGO3}
A.~A. Abramovici {\it et~al.}, Science {\bf 256},  325  (1992).

\bibitem{Bona93}
C. Bona and J. Mass\'o, International Journal of Modern Physics C: Physics and
  Computers {\bf 4},  883  (1993).

\bibitem{Abrahams92a}
A. Abrahams, D. Bernstein, D. Hobill, E. Seidel, and L. Smarr, Phys. Rev.
  D {\bf 45},  3544  (1992).

\bibitem{Anninos93b}
P. Anninos, D. Hobill, E. Seidel, L. Smarr, and W.-M. Suen, Phys. Rev. Lett.
  {\bf 71},  2851  (1993).

\bibitem{Hawking73a}
S.~W. Hawking and G.~F.~R. Ellis, {\em The Large Scale Structure of Spacetime}
  (Cambridge University Press, Cambridge, 1973).

\bibitem{Thorne86}
{\em Black Holes: The Membrane Paradigm}, edited by K.~S. Thorne, R.~H. Price,
  and D.~A. Macdonald (Yale University Press, London, 1986).

\bibitem{Penrose69}
R. Penrose, Riv. Nuovo Cimento {\bf 1},  252  (1969).

\bibitem{Carter79}
B. Carter,  in {\em General Relativity: An Einstein Centenary Survey}, edited
  by S. Hawking and W. Israel (Cambridge University Press, Cambridge, 1979).

\bibitem{Chandrasekhar83}
S. Chandrasekhar, {\em The Mathematical Theory of Black Holes} (Oxford U.
  Press, Oxford, U.K., 1983).

\bibitem{Cook90a}
G. Cook and J. York, Phys. Rev. D {\bf 41}, 1077 (1990).

\bibitem{Hughes94a}
S. Hughes, C.~R.~Keeton II, P. Walker, K. Walsh, S.~L. Shapiro, and S.~A.
  Teukolsky, (1994), preprint.

\bibitem{Bernstein93b}
D. Bernstein, D. Hobill, E. Seidel, J. Towns, and L. Smarr, Phys. Rev. D
  (1994), in preparation.

\bibitem{Anninos93c}
P. Anninos, D. Hobill, E. Seidel, L. Smarr, and J. Towns, {\em Numerical
  Astrophysics} (Springer-Verlag, New York, 1994), to appear.

\bibitem{seidel92a}
E. Seidel and W.-M. Suen, Phys. Rev. Lett. {\bf 69},  1845  (1992).

\bibitem{Anninos94a}
P. Anninos, D. Hobill, E. Seidel, L. Smarr, and W.-M. Suen, Phys. Rev. D
  (1994), in preparation.

\bibitem{Smarr79}
L. Smarr,  in {\em Sources of Gravitational Radiation}, edited by L. Smarr
  (Cambridge University Press, Cambridge, 1979), p.\ 245.

\bibitem{Flanagan91a}
E. Flanagan, Phys. Rev.
  D {\bf 44},  2409  (1991).

\end{thebibliography}
\end{document}